\documentclass[sigconf]{acmart-me}

\usepackage{booktabs} 
\usepackage{url}
\usepackage{color}
\usepackage{enumitem}
\setcopyright{rightsretained}

\acmDOI{}

\acmISBN{}

\acmConference[MediaEval'21]{Multimedia Evaluation Workshop}{December 13-15 2021}{Online} 
\acmYear{2021}
\copyrightyear{}

\acmPrice{}

\begin{document}
\title{Semi-supervised music emotion recognition using noisy student training and harmonic pitch class profiles}

\author{Hao Hao Tan}
\email{helloharry66@gmail.com}

%
%
%
%
%

\renewcommand{\shortauthors}{Hao Hao Tan}
\renewcommand{\shorttitle}{Emotions and Themes in Music}

\begin{abstract}
We present Mirable's submission to the 2021 Emotions and Themes in Music challenge. In this work, we intend to address the question: can we leverage semi-supervised learning techniques on music emotion recognition? With that, we experiment with noisy student training, which has improved model performance in the image classification domain. As the noisy student method requires a strong teacher model, we further delve into the factors including (i) input training length and (ii) complementary music representations to further boost the performance of the teacher model. For (i), we find that models trained with short input length perform better in PR-AUC, whereas those trained with long input length perform better in ROC-AUC. For (ii), we find that using harmonic pitch class profiles (HPCP) consistently improve tagging performance, which suggests that harmonic representation is useful for music emotion tagging. Finally, we find that noisy student method only improves tagging results for the case of long training length. Additionally, we find that ensembling representations trained with different training lengths can improve tagging results significantly, which suggest a possible direction to explore incorporating multiple temporal resolutions in the network architecture for future work.
\end{abstract}

%
%
%
%
%


\maketitle

\section{Introduction}
\label{sec:intro}
Emotions and themes are high-level musical attributes that are abstract and highly subjective. Obtaining emotion labels typically require human annotation, which can be time consuming and potentially costly. Is it possible to use semi-supervised learning techniques, such that we can leverage on unlabelled music tracks to learn emotion tags, while only using a small amount of labelled data?  Following this question, we intend to explore the usage of noisy student training \cite{xie2020self} on music emotion recognition. Recently, \cite{won2021transformer} proposed the music tagging transformer, which also uses noisy student training, but it is applied to general music tagging and does not focus on emotion and theme related tags. Additionally, we explore two other factors to improve the tagging performance of the teacher model: (i) the input training length; (ii) adding music representations to complement the learning of music emotion.

\section{Approach}
\label{sec:approach}
\subsection{Pre-Processing and Augmentation}
We extract Mel-spectrograms with 128 bins from raw audio using a sampling rate of 44.1kHz, and the Mel-spectrograms are down-sampled with an averaging factor of 10 along the temporal dimension. The number of time steps for each Mel-spectrogram vary according to the training strategy, which will be discussed in Section 2.3. For data augmentation, we perform time masking and frequency masking, similar to the idea in SpecAugment \cite{park2019specaugment}. The maximum possible length of both masks vary between 20 to 60, and the value is being sampled randomly for each training batch.

\begin{table}
  \label{tab:freq}
  \begin{tabular}{cccc}
    \toprule
    Model&ROC-AUC&PR-AUC&F-Score\\
    \midrule
    baseline & 0.7258 & 0.1077 & 0.1656\\
    long-normal & 0.7256 & 0.1024 & 0.1578\\
    long-hpcp & 0.7587 & 0.1220 & 0.1854\\
    long-hpcp-noisy & 0.7614 & 0.1235 & 0.1833\\
    short-normal & 0.7477 & 0.1234 & 0.1855\\
    short-hpcp & 0.7541 & 0.1275 & 0.1864\\
    short-hpcp-noisy & 0.7488 & 0.1226 & 0.1804\\
    ensemble & \textbf{0.7687} & \textbf{0.1356} & \textbf{0.1978}\\
  \bottomrule
\end{tabular}
\caption{Test-set performance of our models.}
\vspace{-9mm}
\end{table}

\begin{figure*}
  \includegraphics[width=\textwidth]{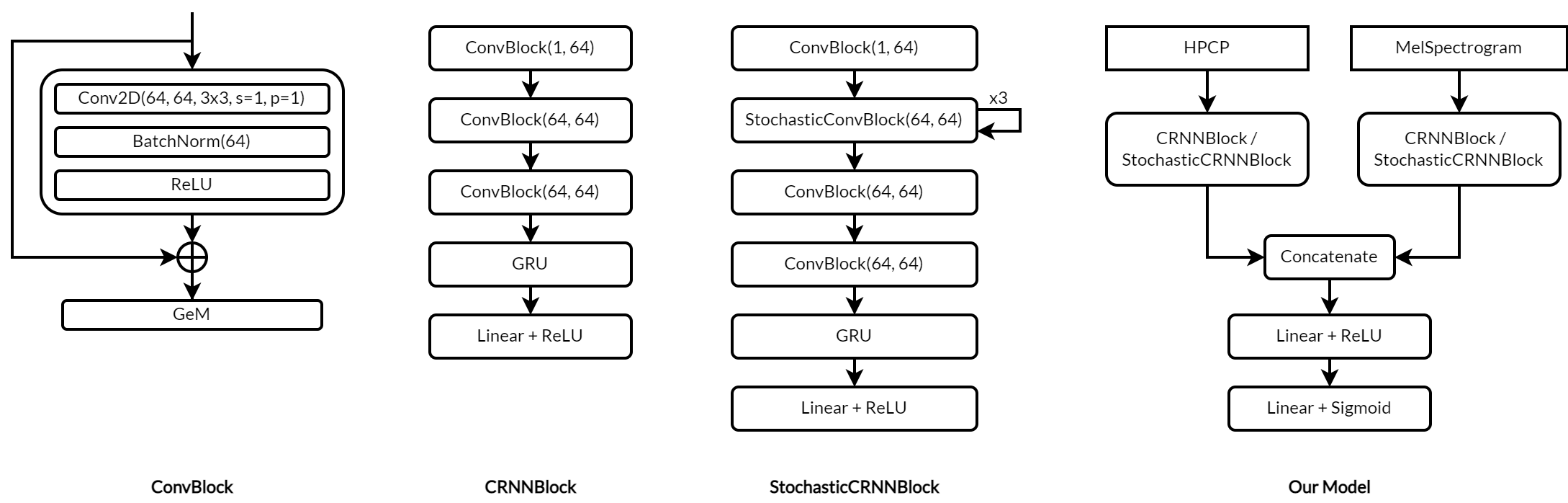}
  \caption{Overview of our model.}
\end{figure*}

\subsection{Model Training}
As shown in Figure 1, our base model architecture is similar to CRNN \cite{choi2017convolutional}, with some revisions which include adding residual connections to our ConvBlock, and using GeMPool \cite{radenovic2018fine} instead of MaxPool. We train all of our models for a maximum of 100 epochs, with an Adam optimizer and learning rate of 0.0001. Early stopping is performed when the validation ROC-AUC does not improve for 5 epochs, and we store the model weights from the epoch with the best ROC-AUC evaluated on the validation set.

\subsection{Long VS Short Training Length}
For the long training length mode, we use the first $\approx$ 185 seconds of the track, which corresponds to 1600 time steps in the Mel-spectrogram after average pooling. For the short training length mode, we chunk each track into samples of length $\approx$ 9.25 seconds, which corresponds to 80 time steps in the Mel-spectrogram after average pooling. During evaluation, we average the logits of all chunks to obtain the final output for each track.

\subsection{Harmonic Pitch Class Profiles (HPCP)}
HPCP \cite{gomez2006tonal} is a type of chroma feature that describes tonality and harmonic content of a music track. We extract HPCP with 12 pitch classes from raw audio using a sampling rate of 44.1kHz. We do not apply average pooling along the temporal dimension of HPCP. The corresponding number of time steps for HPCP are 4000 and 200 for both long and short training length mode respectively. We concatenate the learnt latent features from the Mel-spectrogram and the HPCP block, each with dimension $d=256$, and pass through two linear layers to obtain the fused output.

\begin{table}
  \label{tab:tpr}
  \begin{tabular}{ccc}
    \toprule
    Model&Avg TPR&Avg TNR\\
    \midrule
    long-hpcp-noisy & 0.3645 & \textbf{0.8851}\\
    short-normal & 0.3842 & 0.8737\\
    ensemble & \textbf{0.4099} & 0.8671\\
  \bottomrule
\end{tabular}
\caption{Average true positive rate (TPR) and true negative rate (TNR) for each model across all labels.}
\vspace{-9mm}
\end{table}

\vspace{2mm}
\subsection{Noisy Student Training}
Noisy student training \cite{xie2020self} is an extension of self-training, with the usage of equal-or-larger student models and added noise to improve the representation learnt from the teacher model. To add noise, we enhance data augmentation by increasing the maximum possible masking length to between 30 and 90 for both time and frequency masking, as well as adding standard Gaussian noise with a weight of 0.01. To implement stochastic depth \cite{huang2016deep}, we use 3 StochasticConvBlocks which are ConvBlocks that could be randomly bypassed with a probability of 0.1 each. During evaluation, all the layers will be passed through. StochasticConvBlock also has an additional dropout of probability 0.1 after the ReLU layer. 

In this work, we use the corresponding HPCP models for each long and short training length mode as the teacher model. We only use the predictions which are $> 0.1$ as positive pseudo-labels, and those $< 1e^{-6}$ as negative pseudo-labels. Both decision thresholds are determined by conducting an empirical evaluation on the predicted value distribution using the teacher model, carried out on the training and validation set. We take the leftmost 5\% percentile for the negative label distribution, and the rightmost 5\% percentile for the positive label distribution to ensure better confidence.

\subsection{Model Ensemble}
Finally, we investigate the results of combining the output of both long and short training length models, by simply taking the weighted sum of their best models: $l_{final} = \alpha \cdot l_{short} + (1-\alpha) \cdot l_{long}$. We use the validation set to find the ratio $\alpha$ which gives the best results.

\section{Results and Analysis}

For the training length factor, we find that models trained with long input length perform better in ROC-AUC, but models trained with short input length perform significantly better in PR-AUC. According to Table 2, this is because the former has a higher TNR, while the latter has a higher TPR. Since PR-AUC focuses more on the minority class (in this case the positive class) and ROC-AUC focuses on both, the latter model scores better in PR-AUC. We also find that adding HPCP improves tagging results consistently for both cases, which suggests that harmonic representation is important for music emotion recognition.

For noisy student training, the results are rather inconclusive. We find slight improvements in the long training length case, but the result degrades for the short training length case. Also, we only run noisy student training for 1 iteration, as we find the results consistently degrade for subsequent iterations. Additionally, we try to add more unlabelled tracks from the Lakh MP3 dataset ($\approx 45,000$ $30$ seconds track) to increase the training dataset size, but we do not observe any performance improvement. We infer that noisy student method might not necessarily work well for music emotion recognition tasks, due to the abstract nature and subjectivity of emotion and theme labels. Hence, a small subset of emotion labels might not be sufficient to represent the full dataset. 

For model ensembling, we choose to ensemble the `long-noisy' model and the `short-normal' model. We find that $\alpha=0.7$ is optimal through our validation set, hence suggesting that the final output gives more weightage to the short training length model. From the test set results, we can also see that this ensemble method improves the tagging performance significantly, which suggest that combining different views of audio in terms of temporal resolution can produce better learnt representations. 

\vspace{-2mm}
\section{Discussion and Outlook}
While investigating the related work, we find that this work still uses a relatively long training length (even for short length we use $\approx 9$ seconds, as compared to previous works with $\approx 2$ to $5$ seconds), and low temporal resolution, which we intend to change in our future work. For future work, we are interested in tweaking the network architecture to capture views of different temporal resolutions in the audio sample. We would also like to explore using noisy student training with different model architectures and datasets of a much larger scale.
 
\nocite{bogdanov}
\nocite{bogdanov2}

\bibliographystyle{ACM-Reference-Format}
\def\bibfont{\small} 
\bibliography{test} 

\end{document}